\documentclass[11pt, preprint]{article}
\usepackage{color} \usepackage{cancel}
\usepackage{float} \usepackage{graphicx}
\usepackage{subfigure} \usepackage{caption}	
\usepackage{dcolumn}
\usepackage[toc,page]{appendix} \usepackage{authblk}	 		
\usepackage{indentfirst}	
\usepackage{authblk}	 		
\usepackage{multirow}     
\usepackage{hyperref}     
\usepackage{amsfonts, amssymb, amsmath}
\usepackage[left=2.5cm,right=2.5cm,top=2.9cm,bottom=2.9cm,a4paper]{geometry}
\usepackage[noadjust]{cite}
\usepackage{filecontents}
\usepackage{soul}
\usepackage{xcolor}
\setstcolor{red}

\graphicspath{{./figures/}}
\hypersetup{ colorlinks   = true, 
urlcolor     = blue, 
linkcolor    = blue, 
citecolor    = red 	 
}

\title{\Large\bf Inflation with explicit parametric connection between
General Relativity and Scalar-Tensor gravity}
\author[1]{I. V. Fomin\thanks{ingvor@inbox.ru}}
\author[1,2]{S. V. Chervon\thanks{chervon.sergey@gmail.com}}
\affil[1]{\small \it  Bauman Moscow State Technical University, 2-nd Baumanskaya street, 5, Moscow, 105005, Russia}
\affil[2]{\small \it Ulyanovsk State Pedagogical University, Ulyanovsk, 100-years V.I. Lenin's Birthday Square, B.4, 432071, Russia}

\begin{document}

\maketitle
\begin{abstract}
In this paper we consider the cosmological inflation with scalar-tensor gravity and compare it with standard inflation based on general relativity. The difference is determined by the value of the parameter $\Delta$. This approach is associated with using the special ansatz which links a function that defines a type of gravity with a scale factor of the universe.
\end{abstract}

\section{Introduction}

Despite the fact that the standard theory of cosmological inflation based on General Relativity (GR) solves some puzzles of the early universe and explains the formation of a large-scale structure \cite{Starobinsky:1980te,Guth:1980zm,Linde:1981mu,Albrecht:1982wi,Liddle}, at the present time, according to modern observations of second accelerated expansion of the universe \cite{Perlmutter:1998np,Riess:1998cb},
most of the energy density falls to an unknown component, which is called dark energy.
To explain the nature of dark energy within the framework of GR, the models with the cosmological constant are used, as well as the models of quintessence \cite{Frieman:2008sn}, $K$-essence \cite{Frieman:2008sn, Unnikrishnan:2013vga}, which include some scalar field or several fields \cite{Chervon:2014dya}at the early and modern stages of the universe's evolution.

There are attempts to explain the both accelerated stages by using modified gravity theories \cite{Nojiri:2010wj,Nojiri:2017ncd,Clifton:2011jh,Elizalde:2004mq}, the special case of which is the scalar-tensor gravity (STG) theories \cite{Faraoni,Fujii:2003pa}.

For experimental verification of modified gravity theories in the framework of experiments in Solar System and other astrophysical observations the parameterized post-Newtonian (PPN) formalism is used in which the deviation from GR in the first order is determined by the parameter $\gamma_{PPN}=0.9998\pm0.0003$ \cite{Fomalont:2009zg}, where the value $\gamma_{PPN}=1$ corresponds to Einstein gravity.

In this paper we consider a similar approach at the stage of cosmological inflation in which the difference between models with GR and STG is determined by some parameter $\Delta$ on the level of cosmological perturbations.

For this purpose we postulate the special relationship between function $F(\phi)$ which defines the non-minimal coupling of a scalar field with curvature and compare the models in framework of STG and GR on the basis of the experimental data on CMB anisotropy \cite{Ade:2015xua}.

\section{Friedmann cosmology from GR and STG}

The generalized scalar-tensor theory described by the action~\cite{Faraoni,Fujii:2003pa}
\begin{eqnarray}\label{actionFR}
S_{(STG)} = \int d^4x\sqrt{-g}\Big[\frac{1}{2}F(\phi) R -
 \frac{\omega(\phi)}{2}g^{\mu\nu}\partial_{\mu}\phi \partial_{\nu} \phi
 - V(\phi)\Big] +S^{(m)},\\
S^{(m)}=\int d^4x\sqrt{-g} {\cal L}^{(m)},
\end{eqnarray}
where Einstein gravitational constant  $ \kappa=1$, $g$ is the determinant of space-time metric tensor $g_{\mu\nu}$, $\phi$ is a scalar field with the potential $V=V(\phi)$, $\omega(\phi)$ and $F(\phi)$ are the differentiable functions of $\phi$ which define the type of gravity, $R$ is the Ricci scalar of the space-time, ${\cal L}^{(m)}$ the Lagrangian of the matter.
In the present article we will study the case of vacuum solutions for the model (\ref{actionFR}) suggesting that $ S^{(m)}=0$.

The action of the scalar field Einstein gravity is
\begin{equation}\label{actionR}
S_{(GR)} = \int d^4x\sqrt{-g}\left[\frac{R}{2} -
\frac{1}{2}g^{\mu\nu}\partial_{\mu}\phi \partial_{\nu} \phi-V(\phi)\right].
\end{equation}

Let us remind that the scalar field $\phi$ is a gravitational field in the action (\ref{actionFR}), the scalar field $\phi$ in the action (\ref{actionR}) is the source of the gravitation and it is non-gravitational (material) field.

Considering the gravitational and field equations of the models (\ref{actionFR}) and (\ref{actionR}) in the Friedman universe we use the choice of natural units, including $ \kappa=1$. Such choice formally means that we could not make difference between gravitational and non-gravitational (material) scalar fields. Thus, the solutions of the model's equations should be considered in the subsequent situation. That is, from the physical point of view, the solutions will correspond to different representation of gravity.

Let as note also that the cosmological constant $\Lambda$ can be extracted from the constant part of the potential $V(\phi)$, therefore we did not include it into the actions (\ref{actionFR}) and (\ref{actionR}).

To describe a homogeneous and isotropic universe we chose the Friedmann-Robertson-Walker (FRW) metric in the form
\begin{equation}\label{FRW}
ds^2=-dt^2+a^2(t)\left(\frac{d r^2}{1-k r^2}+r^2 \left( d\theta^2+\sin^2\theta d\varphi^2\right)\right),
\end{equation}
where $a(t)$ is a scale factor, a constant  $k$ is the indicator of universe's
type:
$ k>0,~k=0,~k<0 $ are associated with closed, spatially flat, open universes, correspondingly.

The cosmological dynamic equations for the STG theory (\ref{actionFR}) in a spatially-flat FRW metric are~\cite{DeFelice:2011zh,DeFelice:2011jm}
\begin{eqnarray}
\label{E1}
&& E_{1}\equiv3FH^{2}+3H\dot{F}-\frac{\omega}{2}\dot{\phi}^{2}-V(\phi)=0,\\
\label{E2}
&& E_{2}\equiv3FH^{2}+2H\dot{F}+2F\dot{H}+\ddot{F}+\frac{\omega}{2}\dot{\phi}^{2}-V(\phi)=0,\\
\label{E3}
&&E_{2}\equiv\omega\ddot{\phi} + 3\omega H\dot{\phi}
+\frac{1}{2}\dot{\phi}^{2}\omega'_{\phi}+V'_{\phi}-6H^{2}F'_{\phi}-3\dot{H}F'_{\phi}= 0,
\end{eqnarray}
where a dot represents a derivative with respect to the cosmic time $t$, $H \equiv \dot{a}/a$ denotes the Hubble parameter and $F'_{\phi} = \partial F/\partial \phi $.

From Bianchi identities one has
\begin{equation}
\label{Bianchi}
\dot{\phi}E_{3}+\dot{E}_{1}+3H(E_{1}-E_{2})=0,
\end{equation}
thus, only two of the equations (\ref{E1})--(\ref{E3}) are independent.

For this reason, the scalar field equation (\ref{E3}) can be derived from the equations (\ref{E1})--(\ref{E2}) and equations (\ref{E1})--(\ref{E2}) completely describe the cosmological dynamics and we will deal with the STG gravitational equations only
\begin{equation}
\label{EFR1}
3FH^{2}+3H\dot{F}=\frac{\omega(\phi)}{2}\dot{\phi}^{2}+V(\phi),
\end{equation}
\begin{equation}
\label{EFR2}
H\dot{F}-2F\dot{H}-\ddot{F}=\omega(\phi)\dot{\phi}^{2}.
\end{equation}

We will refer to equations (\ref{EFR1})--(\ref{EFR2}) as for {\it STG cosmology} equations.

If $F=1$ equations (\ref{EFR1})--(\ref{EFR2}) are reduced to those for scalar field Friedmann (inflationary) cosmology from GR
\begin{equation}
\label{ER1}
3H^{2}=\frac{\omega(\phi)}{2}\dot{\phi}^{2}+V(\phi),
\end{equation}
\begin{equation}
\label{ER2}
\omega(\phi)\dot{\phi}^{2}=-2\dot{H},
\end{equation}
where we can chose $\omega=1$ or redefine a scalar field as $\psi=\int\sqrt{\omega(\phi)}d\phi$.

Here, we remember, the scalar field $ \phi$ is the source of Einstein gravity. We will refer to equations (\ref{ER1})-(\ref{ER2}) as for {\it GR cosmology} equations\footnote{We introduced the terms {\it STG cosmology} and {\it GR cosmology} with the aim to distinguish the gravitational theories background. In both cases we deal with Friedman cosmology because of FRW metric of the space-time is applied.}.

The usual way to solve  STG cosmology equations for inflation is the conformal transformation $\hat{g}^{\mu\nu}=F(\phi)g^{\mu\nu}$ from action (\ref{actionFR}) to (\ref{actionR}) (or from Jordan frame to Einstein frame)~\cite{DeFelice:2011zh,DeFelice:2011jm,DeFelice:2011bh,Chakraborty:2016ydo}
\begin{equation}\label{actionR1}
\hat{S} = \int d^4x\sqrt{-\hat{g}}\left[\frac{\hat{R}}{2} -
 \frac{1}{2}\hat{g}^{\mu\nu}\partial_{\mu}\hat{\phi} \partial_{\nu} \hat{\phi}
 - \hat{V}(\hat{\phi})\right].
\end{equation}
where
\begin{equation}
\label{TR1}
\hat{V}(\hat{\phi})=V(\phi)/F^{2},~~\hat{\phi}=\int\sqrt{\frac{3}{2}\left(\frac{F'}{F}\right)^{2}+\frac{\omega}{F}}\, d\phi
\end{equation}
and the connections between the variables in the two frames are
\begin{equation}
\label{TR2}
d\hat{t}=\sqrt{F}dt,~~\hat{a}=\sqrt{F}a,~~\hat{H}=\frac{1}{\sqrt{F}}\left(H+\frac{\dot{F}}{2F}\right).
\end{equation}

For the action (\ref{actionR1}) in the spatially flat Friedmann universe we have  GR cosmology equations (\ref{ER1})--(\ref{ER2}) in terms of $\hat{\phi}, \hat{H}, \hat{V}(\hat{\phi})$, $\hat{t}$ and, also, with $\hat{\omega}=1$.
Thus, knowing the solutions in GR cosmology, one can find corresponding ones in STG cosmology using inverse to (\ref{TR1})--(\ref{TR2}) transformations.

In this article we propose the new method to construct the exact solutions of equations (\ref{EFR1})--(\ref{EFR2}) by the special choice of the functions $F(\phi)$ and $\omega(\phi)$ and to compare the parameters of cosmological perturbations with ones in GR cosmology.

\section{Cosmological perturbations}\label{Cosmological perturbations}
For calculating of the parameters of cosmological perturbations in linear order we will use for STG cosmology the known formulas obtained in the papers~\cite{DeFelice:2011zh,DeFelice:2011jm,DeFelice:2011bh}.

Firstly, we write the functions
\begin{eqnarray}
\label{w1}
&& w_{1}=F,\\
\label{w2}
&& w_{2}=2HF+\dot{F},\\
\label{w3}
&& w_{3}=-9FH^{2}-9H\dot{F}+\frac{3}{2}\omega(\phi)\dot{\phi}^{2},\\
\label{w4}
&& w_{4}=F.
\end{eqnarray}

The power spectrum of the curvature perturbation is given by~\cite{DeFelice:2011zh,DeFelice:2011jm,DeFelice:2011bh}
\begin{equation}
\label{PR}
{\cal P}_{{\rm S}}=\frac{H^{2}}{8\pi^{2}Q_{S}c_{S}^{3}},
\end{equation}
where
\begin{eqnarray}
\label{Qs}
Q_{S} & \equiv & \frac{w_{1}(4w_{1}w_{3}+9w_{2}^{2})}{3w_{2}^{2}},\\
\label{cs}
c_{S}^{2} & \equiv & \frac{3(2w_{1}^{2}w_{2}H-w_{2}^{2}w_{4}
+4w_{1}\dot{w}_{1}w_{2}-2w_{1}^{2}\dot{w}_{2})}
{w_1(4w_{1}w_{3}+9w_{2}^{2})}.
\end{eqnarray}
Here $c_{S}$ is the velocity of the scalar perturbations.

If $c_{S}$ is a constant, we have $d\ln k$ at $c_{S}k=aH$ may be written as  $d\ln k=H+\frac{\dot{H}}{H}dt=H(1-\epsilon)dt$.

In this case we have the scalar spectral index
\begin{eqnarray}
\label{nR}
n_{{\rm S}}-1\equiv\frac{d\ln{\cal P}_{{\rm S}}}{d\ln k}\bigg|_{c_{S}k=aH}=\frac{{\cal \dot{P}}_{{\rm S}}}{H(1-\epsilon){\cal P}_{{\rm S}}}\bigg|_{c_{S}k=aH}~~,
\end{eqnarray}
where $\epsilon=-\frac{\dot{H}}{H^{2}}=1-\frac{\ddot{a}a}{\dot{a}^{2}}$ is the slow-roll parameter.

The tensor power spectrum is given by \cite{DeFelice:2011zh,DeFelice:2011jm,DeFelice:2011bh}
\begin{equation}
\label{Pt}
{\cal P}_{{\rm T}}=\frac{H^{2}}{2\pi^{2}Q_{T}c_{T}^{3}}
\end{equation}
where
\begin{eqnarray}
\label{Qt}
Q_{T} & \equiv &\frac{ w_{1}}{4},\\
\label{ct}
c_{T}^{2} & \equiv & \frac{w_{4}}{w_{1}}.
\end{eqnarray}

The spectral index of tensor perturbations is
\begin{eqnarray}
\label{nt}
n_{{\rm T}}\equiv\frac{d\ln{\cal P}_{{\rm T}}}{d\ln k}\bigg|_{c_{T}k=aH}=\frac{{\cal \dot{P}}_{{\rm T}}}{H(1-\epsilon){\cal P}_{{\rm T}}}\bigg|_{c_{T}k=aH}~~.
\end{eqnarray}

It needs to mention that the spectral indices (\ref{nR}) and (\ref{nt}) differ from ones given in works \cite{DeFelice:2011zh,DeFelice:2011jm,DeFelice:2011bh} by the factor $1/(1-\epsilon)$ since we don't use the slow-roll approximation at this step of calculations. In the case of slow-roll approximation we have $1-\epsilon\approx 1$.

Tensor-to-scalar ratio is
\begin{equation}
\label{tsr}
r=\frac{{\cal P}_{{\rm T}}}{{\cal P}_{{\rm S}}}=4\frac{Q_{S}}{Q_{T}}\left(\frac{c_{S}}{c_{T}}\right)^{3}.
\end{equation}

Also, we note that direct calculations by the above formulas give the following propagation velocities of cosmological perturbations $c_{S}=1$ and $c_{T}=1$ for any type of scalar-tensor gravity \cite{Pozdeeva:2016cja}, i.e. similar as in the case of standard GR cosmology.

The parameters of cosmological perturbations are the basis for verification of cosmological models from the observational data on CMB anisotropy \cite{Ade:2015xua} in the inflationary paradigm of early universe.

\section{The specific connection of the coupling function $F(\phi)$ with the scale factor}\label{scale factor}
In this section we will consider the connection of the function $F(t)\equiv F(\phi(t))$ with the scale factor $a(t)$ in the following form
\begin{equation}
\label{ansatz}
F(t)=1-\frac{\gamma}{a^{2}(t)}=1-\Delta(t),
\end{equation}
where $\gamma$ is the constant and $\Delta=\gamma/a^{2}(t)$ is the dimensionless parameter which defines the deviation of function $F$ from unit.

The equations  (\ref{EFR1}) -- (\ref{EFR2}) are reduced to
\begin{equation}
\label{EFR3}
V(\phi)=3H^{2}+\dot{H},
\end{equation}
\begin{equation}
\label{EFR4}
\frac{\omega(\phi)}{2}\dot{\phi}^{2}=-\dot{H}+3\gamma\left(\frac{H}{a}\right)^{2}.
\end{equation}

From the expressions for the parameters of cosmological perturbations which were considered in Sec. \ref{Cosmological perturbations}, using the equations (\ref{ansatz}) -- (\ref{EFR4}), we obtain
\begin{equation}
c_{S}=1,~~~c_{T}=1,
\end{equation}
\begin{eqnarray}
\label{r}
\nonumber
r_{(STG)}=16\left(1-\frac{\ddot{a}a}{\dot{a}^{2}}+\frac{\gamma\ddot{a}}{a\dot{a}^{2}}+\frac{2\gamma}{a^{2}}\right)=\\
16\left[\epsilon +\frac{\gamma}{a^{2}}(3-\epsilon)\right]=16[\epsilon(1-\Delta) +3\Delta)]\,,
\end{eqnarray}
\begin{equation}
\label{PS}
{\cal P}_{{\rm S}(STG)}=\frac{2\dot{a}^{2}}{\pi^{2}(a^{2}-\gamma)r}=\frac{2H^{2}}{\pi^{2}(1-\Delta)r}\,,
\end{equation}
\begin{equation}
\label{PT}
{\cal P}_{{\rm T}(STG)}=\frac{2\dot{a}^{2}}{\pi^{2}(a^{2}-\gamma)}=\frac{2H^{2}}{\pi^{2}(1-\Delta)}\,,
\end{equation}
\begin{eqnarray}
\label{nS}
n_{{\rm S}(STG)}-1=\frac{1}{1-\epsilon}\left[-2\epsilon-\frac{2\gamma}{a^{2}-\gamma}-\frac{\dot{r}}{Hr}\right]=
\frac{1}{1-\epsilon}\left[-2\epsilon-\frac{\dot{r}}{Hr}-\frac{2\Delta}{1-\Delta}\right],
\end{eqnarray}
\begin{equation}
\label{nT}
n_{{\rm T}(STG)}=\frac{1}{1-\epsilon}\left[-2\epsilon-\frac{2\gamma}{a^{2}-\gamma}\right]=
\frac{1}{1-\epsilon}\left[-2\epsilon-\frac{2\Delta}{1-\Delta}\right].
\end{equation}

 The parameters of cosmological perturbations are calculated on the crossing of Hubble radius $(k=aH)$ with corresponding time $t_{H}\approx t_{e}$ \cite{DeFelice:2011zh,DeFelice:2011jm,DeFelice:2011bh}, where $t_{e}$ is the time of ending of the inflation, also, $\Delta<1$.

Thus, the parameters of cosmological perturbations differ from ones for standard inflation and the difference is determined by the constant $\gamma$ or by means of the parameter $\Delta$.

In the case of $\gamma=0$ or $\Delta=0$ one has the parameters of cosmological perturbations for standard inflation \cite{Liddle,Chervon:2008zz,Fomin:2017xlx}
\begin{equation}
\label{rGR}
r_{(GR)}=16\epsilon,~~~{\cal P}_{{\rm S}(GR)}=\frac{H^{2}}{8\pi^{2}\epsilon},~~~{\cal P}_{{\rm T}(GR)}=\frac{2H^{2}}{\pi^{2}},
\end{equation}
\begin{equation}
\label{nGR}
n_{{\rm S}(GR)}-1=-\frac{2\epsilon+\frac{\dot{\epsilon}}{H\epsilon}}{1-\epsilon}=2\left(\frac{\delta-2\epsilon}{1-\epsilon}\right),~~~
n_{{\rm T}(GR)}=-\frac{2\epsilon}{1-\epsilon},
\end{equation}
\begin{equation}
\label{delta}
\delta=\epsilon-\frac{\dot{\epsilon}}{2H\epsilon}=-\frac{\ddot{H}}{2H\dot{H}}.
\end{equation}

Also, when $\gamma=0$, the dynamical equations (\ref{EFR3})--(\ref{EFR4}) are reduced to the equations for GR cosmology (\ref{ER1})--(\ref{ER2}).

The difference between parameters of cosmological perturbations for inflation in STG cosmology (\ref{r})--(\ref{nT}) and GR one (\ref{rGR})--(\ref{nGR}) is
\begin{equation}
\label{R}
\frac{r_{(STG)}}{r_{(GR)}}=1-\Delta+3\frac{\Delta}{\epsilon}
\end{equation}
\begin{equation}
\label{PSTG}
\frac{{\cal P}_{{\rm S}(STG)}}{{\cal P}_{{\rm S}(GR)}}=\frac{{\cal P}_{{\rm T}(STG)}}{{\cal P}_{{\rm T}(GR)}}=\frac{1}{1-\Delta}
\end{equation}
\begin{equation}
\label{NT}
n_{{\rm T}(STG)}-n_{{\rm T}(GR)}=-\frac{2\Delta}{(1-\epsilon)(1-\Delta)}
\end{equation}
\begin{eqnarray}
\label{NS}
n_{{\rm S}(STG)}-n_{{\rm S}(GR)}=-\frac{1}{1-\epsilon}\left[\frac{\dot{\Delta}(3-\epsilon)}{[\Delta(3-\epsilon)+\epsilon]H}\nonumber
+\frac{6\Delta(\delta-\epsilon)}{\Delta(3-\epsilon)+\epsilon}+\frac{2\Delta}{1-\Delta}\right]=\\ \nonumber
-\frac{1}{1-\epsilon}\left[\frac{6\Delta(\delta-\epsilon)-2\Delta(3-\epsilon)}{\Delta(3-\epsilon)+
\epsilon}+\frac{2\Delta}{1-\Delta}\right]=\\
-\frac{2\Delta}{1-\epsilon}\left[\frac{3\delta-2\epsilon-3}{\Delta(3-\epsilon)+\epsilon}+\frac{1}{1-\Delta}\right],
\end{eqnarray}
where we use the equation (\ref{delta}) and expression $\dot{\Delta}/\Delta=-2H$.

The observational constraints on the parameters of cosmological perturbations from PLANCK \cite{Ade:2015xua} are
\begin{equation}
\label{PLANCK1}
10^{9}{\mathcal{P}}_{S}=2.142\pm0.049, ~~n_{S}=0.9667\pm0.0040,
\end{equation}
\begin{equation}
\label{PLANCK2}
{\mathcal{P}}_{T}=r{\mathcal{P}}_{S},~~r< 0.112.
\end{equation}

Now we estimate the value of parameter $\Delta$ for  power-law inflation with Hubble parameter $H=m/t$ from the observational constraints (\ref{PLANCK1})--(\ref{PLANCK2}). The parameters of cosmological perturbations for standard power-law inflation on the basis of equations (\ref{rGR})--(\ref{delta}) were considered earlier in papers \cite{Chervon:2005zz,Chervon:2008zz,Fomin:2017xlx}.

From equations (\ref{rGR})--(\ref{delta}) we have
\begin{equation}
\label{PL}
r_{(GR)}=\frac{16}{m},~~~ n_{{\rm S}(GR)}-1=n_{{\rm T}(GR)}=\frac{2}{1-m}.
\end{equation}

Further, we consider the following values of spectral tilt of scalar perturbations: $0.97\leq n_{{\rm S}(GR)}\leq0.96$, which correspond to (\ref{PLANCK1}).

For this values we have
\begin{equation}
\label{PlGR}
68\leq m\leq51,~~~-0.03\leq n_{{\rm T}(GR)}\leq-0.04,~~~~0.23\leq r_{(GR)}\leq0.3.
\end{equation}

As one can see, the values of $r_{(GR)}$ don't correspond to the observational constraint (\ref{PLANCK2}).

Now, we chose the value $r_{(STG)}=0.1$, which corresponds to (\ref{PLANCK2}) to estimate the maximum value of $\Delta$ or maximum deviation from GR. For this value of $r_{(STG)}$, from equations (\ref{R})--(\ref{NS}), we have
\begin{eqnarray}
\label{PlSTG}
-0.004\leq \Delta\leq-0.003,~0.976\leq n_{{\rm S}(STG)}\leq0.968,\\
-0.024\leq n_{{\rm T}(STG)}\leq-0.030.
\end{eqnarray}

Thus, for the case of power-law inflation, maximum deviation from GR on the crossing of Hubble radius is order of $10^{-3}$ and ${\cal P}_{{\rm S}(STG)}\simeq{\cal P}_{{\rm S}(GR)}$, ${\cal P}_{{\rm T}(STG)}\simeq{\cal P}_{{\rm T}(GR)}$. Also, from equation (\ref{ansatz}), we have $\Delta(t)=\frac{\gamma}{a^{2}_{0}}t^{-2m}$.

Also we note, that the "resurrecting" procedure for power-law inflation by the value of the tensor-to-scalar ratio on the basis of K-essence model with GR was considered in the paper \cite{Unnikrishnan:2013vga}.

\section{The exact solutions of dynamical equations}
For confrontation of STG inflation with GR inflation we will use the special choice of kinetic function $\omega(\phi)$ which leads to two equivalent systems of equations from (\ref{EFR3})--(\ref{EFR4})
\begin{equation}
\label{EFR5}
F(t)=1-\frac{\gamma}{a^{2}(t)},~~F(\phi)=1-\frac{\gamma}{a^{2}(\phi)},
\end{equation}
\begin{equation}
\label{EFR6}
\omega(t)=1-3\gamma\frac{H^{2}}{\dot{H}a^{2}},~~~
\omega(\phi)=1+3\gamma\left(\frac{H}{aH'_{\phi}}\right)^{2},
\end{equation}
\begin{equation}
\label{EFR7}
V(t)=3H^{2}+\dot{H},~~~V(\phi)=3H^{2}-2H'^{2}_{\phi},
\end{equation}
\begin{equation}
\label{EFR8}
\dot{\phi}^{2}=-2\dot{H},~~~~~\dot{\phi}=-2H'_{\phi}.
\end{equation}

One can find the scale factor $a=a(\phi)$ from Hubble parameter $H=H(\phi)$ by using the equation
\begin{equation}
\label{stFpha3}
a(\phi)=a_{0}\exp\left(-\frac{1}{2}\int\frac{H}{H'_{\phi}}d\phi\right),    \,\,\,\,\,\ -2\left(\frac{a'_{\phi}}{a}\right)=\frac{H}{H'_{\phi}}.
\end{equation}

The exact solutions of equations (\ref{EFR7})--(\ref{EFR8}) for some models of standard inflation and classification of the methods for their generation one can find in the papers \cite{Fomin:2017xlx,Chervon:2017kgn}.

Now, we consider the case of power-law inflation $H=m/t$, the exact solutions from equations (\ref{EFR7})--(\ref{EFR8}) are
\begin{equation}
\phi(t)=\sqrt{2m}\ln t,
\end{equation}
\begin{equation}
V(\phi)=m(3m-1)\exp\left(-{\sqrt{\frac{2}{m}}}\phi\right),
\end{equation}
\begin{equation}
\omega(\phi)=1+\frac{3m\gamma}{a^{2}_{0}}\exp\left(-\sqrt{2}\phi\right),
\end{equation}
\begin{equation}
F(\phi)=1-\frac{\gamma}{a^{2}_{0}}\exp\left(-\sqrt{2}\phi\right).
\end{equation}

Thus, one can use the exact solutions for various models of standard cosmological inflation and reconstruct the functions $F(\phi)$ and $\omega(\phi)$, i.e. to determine the type of STG gravity. Also, one can estimate the deviation $\Delta$ on the level of cosmological perturbations through the procedure described above.

\section{Conclusion and discussions}
In this paper we investigate the inflation with special choice of the function defining the type of scalar-tensor gravity $F(t)=1-\gamma/a^{2}(t)$, which allows us to compare it with inflation of GR cosmology.
Also, we propose the procedure for reconstructing the scalar-tensor gravity theory based on the requirement of coincidence of the background solutions for STG and GR cosmology.

The difference between STG and GR inflation was determined by parameter $\Delta=\gamma/a^{2}(t)$ which effects the values of the parameters of cosmological perturbations.

On the basis of the proposed method we find that the maximum deviation from GR for power-law inflation at the end of inflation is the order of $10^{-3}\ll1$.

In general case of arbitrary inflationary models with $\epsilon\ll1$, $\Delta\ll 1$, $3\Delta\approx\Delta$ from equations (\ref{r})--(\ref{nT}), one has
\begin{equation}
\label{rsr}
r\approx16(\epsilon+\Delta),~~{\cal P}_{{\rm S}}\approx\frac{2H^{2}}{\pi^{2}r},~~
{\cal P}_{{\rm T}}\approx\frac{2H^{2}}{\pi^{2}}
\end{equation}
\begin{equation}
\label{nSsr}
n_{{\rm S}}-1\approx-2(\epsilon+\Delta)-\frac{\dot{r}}{Hr},
~~~~~~n_{{\rm T}}\approx-2(\epsilon+\Delta)
\end{equation}

Therefore, we have the same formulas as for standard inflation~\cite{Liddle} in the first order of slow-roll approximation with shifted slow-roll parameters $\varepsilon=\epsilon+\Delta$ and $\sigma=\varepsilon-\frac{\dot{\varepsilon}}{2H\varepsilon}$
\begin{equation}
\label{rsr1}
r\approx16\varepsilon,~~{\cal P}_{{\rm S}}\approx\frac{H^{2}}{8\pi^{2}\varepsilon},~~
{\cal P}_{{\rm T}}\approx\frac{2H^{2}}{\pi^{2}}
\end{equation}
\begin{equation}
\label{nSsr1}
n_{{\rm S}}\approx1-4\varepsilon+2\sigma,
~~~~~~n_{{\rm T}}\approx-2\varepsilon
\end{equation}

Thus, on the basis of expressions (\ref{rsr1})--(\ref{nSsr1}) one can consider the parameters of cosmological perturbations for STG inflation and compare them with ones for standard inflation in the first order of slow-roll approximation.
The analysis of standard inflationary models in the slow-roll approximation one can find, for example, in the review \cite{Martin:2013tda}.

\section*{Acknowledgments}

I.V. Fomin was supported by RFBR grants 16-02-00488 A and 16-08-00618 A.

\end{document}